\documentclass[twocolumn,amsmath,amssymb]{revtex4}

\usepackage{float}
\usepackage{graphicx}
\usepackage{hyperref}
\hypersetup{colorlinks=true}
\usepackage{wasysym}
\usepackage{algpseudocode}
\usepackage{algorithm}
\begin{document}

\title{An $O(N^3)$ algorithm for the calculation of far field radiation patterns}

\date{\today}

\author{Travis M. Garrett}
\email{tgarrett@stellarscience.com}
\affiliation{Stellar Science}

\begin{abstract}

We present a new method for computing the Near-To-Far-Field (NTFF) transformation  
in FDTD simulations which has an overall scaling of $O(N^3)$  instead of the 
standard $O(N^4)$. 
By mapping the far field with a cartesian coordinate system the 
2D surface integral can be split into two successive 1D integrals.
For a near field spanned by $N_X \times N_Y$ 
discrete sample points, and a far field spanned by $N_{\theta} \times N_{\phi}$ points, 
the calculation can then be performed in $O(N_{\theta}N_X N_Y) + O(N_{\theta}N_{\phi}N_Y)$ operations  
instead of $O(N_{\theta}N_{\phi}N_X N_Y)$.

\end{abstract}

\maketitle

\newcommand{\be}{\begin{equation}}
\newcommand{\ee}{\end{equation}}

\section{Introduction}

The E\&M fields surrounding various real world objects are often highly complex, 
thus motivating their numerical simulation when they need to be accurately understood.
At large distances from the object of interest the fields generally become somewhat simpler:
the near fields which fall off as $1/r^2$ or faster for a radial separation $r$ 
have negligible amplitudes, leaving only the far fields: $1/r$ envelope 
electromagnetic waves which radiate energy to infinity \cite{Jackson_book}.
Accurate evaluations of these far field patterns are needed for a large class of simulations, 
from the classic cases of antennas and radar cross sections, 
to more recent examples such as the backscattering of visible light from biological tissues 
\cite{Drezek_1999,Li_2004,Li_2004_2}.

The Finite Difference Time Domain (FD-TD) method \cite{Taflove_book, Taflove_2007} 
encompasses a powerful class of numerical techniques which enable these simulations.
A particular computational model may 
make use of the Yee scheme to finite differencing the E\&M fields \cite{Yee_1966}, 
specializing to Dey-Mittra methods next to curved bodies \cite{Dey_1997, Yu_2000}, 
using Auxiliary Differential Equations for dispersive materials \cite{Okoniewski_1997}, 
and Berenger type absorbing boundary conditions 
on the exterior of the domain \cite{Berenger_1994, Roden_2000}.
To calculate the far field radiation patterns in these FD-TD simulations
 it turns out that it is not necessary to extend the 
volume of the computational domain out into the far field zone 
(note that we generally assume 3D simulations here).
In the Near-To-Far-Field (NTFF) transformation developed by Umashankar and Taflove \cite{Umashankar_1982,Taflove_1983} 
one instead records the E\&M fields on a 2 dimensional surface near the object, 
transforms this time series data into the frequency domain via a FFT,
converts the resulting phasor fields into effective surface currents for a chosen frequency, 
and then integrates these currents 
using a Green's function to extract the $1/r$ scaling far fields.

If one is interested in the full bistatic radiation pattern across the 2-sphere at infinity 
then this NTFF method traditionally corresponds to roughly an order $O(N^4)$ calculation 
(for each frequency of interest): for each of $N^2$ sample points on the 2-sphere one needs to  
perform an order $O(N^2)$ integration of the surface currents.
Note that we are being somewhat loose with our definitions here, as these $N$s refer to 
different quantities. More precisely, the sample points on the 2-sphere could correspond 
to discretizing the standard $\theta$ and $\phi$ spherical coordinates 
(we use the physics convention, e.g. $N_{\theta} = 180$ and 
$N_{\phi} = 360$ for 1 degree spacing), although one could alternatively map the 2-sphere with, 
say, a recursive triangulation of an icosahedron (see e.g. \cite{Thurston_1998}).
In turn the $N$s for the 2D surface integrals stem from the discretization of the FD-TD domain: 
with $N_X N_Y$ integration points for each of the two $X$-$Y$ planes, 
and likewise $N_X N_Z$ and $N_Y N_Z$ points for the respective $X$-$Z$ and $Y$-$Z$ planes.
If we simplify a bit again and just consider integrating currents on the $X$-$Y$ plane, 
then the overall scaling of the 
NTFF method goes as $O(N_{\theta} N_{\phi} N_X N_Y)$.

We report here a substantial improvement upon this rough $O(N^4)$ scaling.
The key is to map the 2-sphere at infinity with a cartesian coordinate system. 
In this case the 2D surface integral becomes separable, and can be split into 
two successive 1D integrations.
These two 1D integrations are linked through a new temporary variable $\bold{T}$, 
which requires 2D storage (e.g. $\sim N_{\theta} N_Y$ bytes).
This splitting allows the new NTFF algorithm to scale as:
 $O(N_{\theta}N_X N_Y) + O(N_{\theta}N_{\phi}N_Y)$, or $O(N^3)$ roughly speaking.
 
 For completeness we note that in general one is not forced to scale all 
 of the $N_{\theta}, N_{\phi}, N_X, N_Y$ variables up simultaneously: 
 for instance one could scale up just $N_X, N_Y$ by doing successively higher resolution 
 FD-TD runs, while holding the $N_{\theta}, N_{\phi}$ sample points constant.
 In this case both the ``$O(N^4)$" and ``$O(N^3)$" methods would 
 strictly speaking have $O(N^2)$ scaling, although the new $O(N^3)$ algorithm will 
 have a much smaller leading coefficient. That said, as one progresses to simulations 
 of systems with increasing electrical size it is often the case that finer lobes will be produced, 
 thus necessitating increasing the angular resolution of the far fields 
 along with finer resolution in the near fields.
 In practical terms the 
 new method provides a very large speedup.
 For medium sized FDTD problems, with 
 $N_X, N_Y$, and $N_Z$ having values in the hundreds, 
 and 1 degree far field spacing ($N_{\theta}=180, N_{\phi}=360$), 
 we find the new algorithm is over 100 times faster, and 
 expect to find even greater speedups for large problems which run on clusters.
 
 
\section{The $O(N^3)$ NTFF Algorithm}
We begin with a quick overview of the NTFF algorithm, and transition to the details 
needed to implement the $O(N^3)$ version of it. As noted, a simulation doesn't need to extend 
to the radiation zone in order to extract the far fields.
Instead, using Huygen's principle it suffices to record the electric and magnetic fields 
as a function of time on a 2D surface $S$ (mapped by $\bold{r}'$) just outside of the simulated object.
The time series data for these fields can be converted into the frequency domain via a 
Fast Fourier Transform (FFT), and these phasor fields $\bold{E}$ and $\bold{H}$ 
can be subsequently converted into 
effective electric $\bold{J}_s$ and magnetic $\bold{M}_s$ surface currents \cite{Schelkunoff_1951}:
\be
\bold{J}_s (\bold{r}')  = \hat{n}' \times \bold{H} (\bold{r}') ,
\ee
\be
\bold{M}_s (\bold{r}')  = - \hat{n}' \times \bold{E} (\bold{r}')  ,
\ee
where $\hat{n}'$ is the normal to the surface.
These currents can then be integrated over $S$ to determine 
the vector potential fields $\bold{A}$ and $\bold{F}$ at any location $\bold{r}$:
\be
\bold{A}(\bold{r}) = \oiint_{S'} \bold{J}_s(r') g(\bold{r},\bold{r}') dS' , \label{A_eq}
\ee
\be
\bold{F}(\bold{r}) = \oiint_{S'} \bold{M}_s(r') g(\bold{r},\bold{r}') dS' , \label{F_eq}
\ee
where we use the outgoing wave Green's function $g(\bold{r},\bold{r}')$:
\be
g(\bold{r},\bold{r}') = \frac{e^{ik|\bold{r}-\bold{r}'|}}{4 \pi |\bold{r}-\bold{r}'| } \label{green_fn}
\ee

The $\bold{E}$ and $\bold{H}$ fields at any position can in turn be derived from
the vector potentials $\bold{A}$ and $\bold{F}$: 
\be
\bold{E}(\bold{r}) = i\omega \bold{A} - \frac{1}{i\omega \mu \epsilon} \nabla (\nabla \cdot \bold{A}) 
- \frac{1}{\epsilon}\nabla \times \bold{F} \label{E_fromAF}
\ee
\be
\bold{H}(\bold{r}) = \frac{1}{\mu} \nabla \times \bold{A} + i \omega \bold{F} 
- \frac{1}{i\omega \mu \epsilon} \nabla (\nabla \cdot \bold{F}) \label{H_fromAF}
\ee

One can then insert the potentials (\ref{A_eq}) and (\ref{F_eq}) into (\ref{E_fromAF}) and (\ref{H_fromAF}) and expand out, 
expressing $\bold{E}$ and $\bold{H}$ in terms of the currents $\bold{J}_s$ and $\bold{M}_s$.
We only need the far field components of E\&M however (which scale radially as $1/r$) 
and this simplifies the resulting final expressions.
In particular the $|\bold{r}-\bold{r}'|$ terms within the expansions will be simplified in one of two ways, 
depending on whether they occur inside the exponential seen in (\ref{green_fn}) or elsewhere.
Inside of the exponential terms we use:
\be
|\bold{r}-\bold{r}'| = \sqrt{ r^2 + r'^2 - 2 \bold{r} \cdot \bold{r}'} \sim r - r' \cos{\psi} \label{r_phase}
\ee
(with $\cos{\psi} = \bold{r} \cdot \bold{r}' / r r'$,  $r'=|\bold{r}'|$) in order to preserve the phase information. 
This yields:
\be
e^{ik|\bold{r}-\bold{r}'|} = e^{ikr}e^{-ikr'\cos{\psi}} ,
\ee
where the $e^{ikr}$ term does not depend on the $\bold{r}'$ coordinates and 
can be pulled out of the integral.
Everywhere else we simply use $|\bold{r}-\bold{r}'| \sim r$, and drop any terms that 
fall off faster than $1/r$. The resulting equations for $\bold{E}$ and $\bold{H}$ are:

\be
\begin{split}
\bold{E}(\bold{r}) = ik \frac{e^{ikr}}{4\pi r}  \oiint_{S'} 
\Big( \eta [\bold{J}_s -(\bold{J}_s\cdot \hat{r})\hat{r}]  \\
+ [\bold{M}_s \times \hat{r}] \Big) 
e^{ -ik r' \cos{\psi} } dS'
\end{split}
\ee

\be
\begin{split}
\bold{H}(\bold{r}) = ik \frac{e^{ikr}}{4\pi r}  \oiint_{S'} 
\Big( \frac{1}{\eta} [\bold{M}_s -(\bold{M}_s\cdot \hat{r})\hat{r}]  \\
+ [\hat{r} \times \bold{J}_s]   \Big) e^{ -ik r' \cos{\psi} } dS'
\end{split}
\ee
where $k=\omega/c=\omega\sqrt{\mu\epsilon}$ and $\eta=\sqrt{\mu/\epsilon}$.

These can be simplified further by the introduction of 
the mixed field variables $\bold{N}$ and $\bold{L}$ 
which directly integrate the surface currents $\bold{J}_s$ 
and $\bold{M}_s$:

\be
\bold{N}(\theta,\phi) = \oiint_{S'} \bold{J}_s(r') e^{ -ik r' \cos{\psi} } dS' \label{N_eq}
\ee

\be
\bold{L}(\theta,\phi) = \oiint_{S'} \bold{M}_s(r') e^{ -ik r' \cos{\psi} } dS' \label{L_eq}
\ee

$\bold{E}$ and $\bold{H}$ can then be expressed in terms of $\bold{N}$ and 
$\bold{L}$:
\be
E_{\theta}  = ik\frac{e^{ikr}}{4\pi r}\Big( \eta N_{\theta} + L_{\phi} \Big)
, 
E_{\phi}  = ik\frac{e^{ikr}}{4\pi r}\Big( \eta N_{\phi} - L_{\theta} \Big) \label{E_final}
\ee
\be
H_{\theta}  = ik\frac{e^{ikr}}{4\pi r}\Big( - N_{\phi} + \frac{L_{\theta}}{\eta} \Big)
, 
H_{\phi}  = ik\frac{e^{ikr}}{4\pi r}\Big( N_{\theta} + \frac{L_{\phi}}{\eta} \Big) \label{H_final}
\ee
with $E_r = H_r = 0$ since the outgoing waves propagate 
normal to the sphere (note that here $N_{\theta}$ and $N_{\phi}$ are vector 
components of $\bold{N}$, not discretization counts like $N_X$, $N_Y$...).
One can then proceed with further analysis: dividing the local power values 
by the $4\pi$ averaged power to get the directivity, and so forth.

\begin{figure}[h!]
  \centering
    \includegraphics[width=0.6\textwidth]{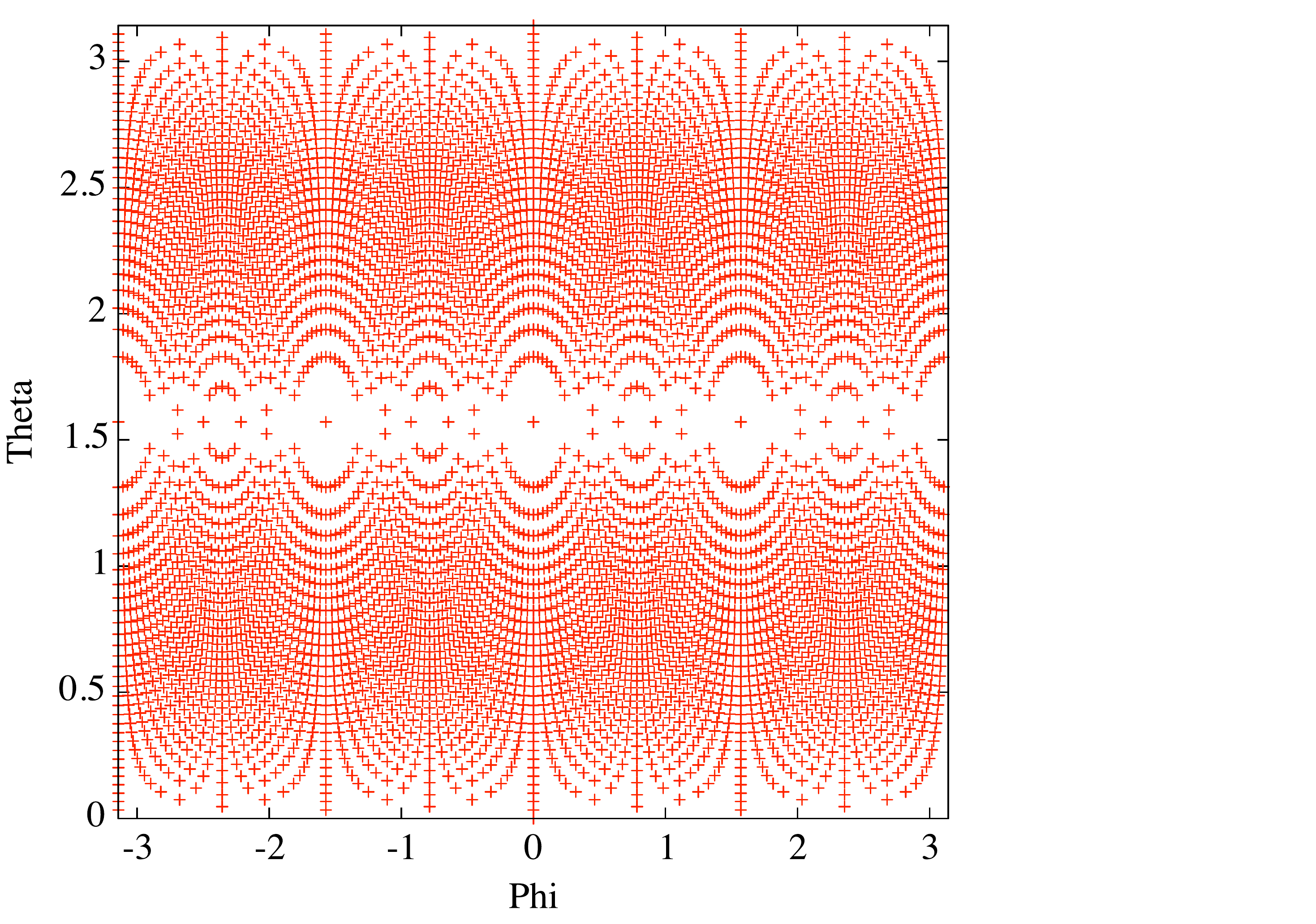}
      \caption{Sampled $(\theta,\phi)$ points on the sphere, 
      given uniform $x$ and $y$ distributions 
      (with $N_{Xfar}=N_{Yfar}=61$).} 
      \label{uniform_xy_plot}
\end{figure}

\begin{figure}[h!]
  \centering
    \includegraphics[width=0.6\textwidth]{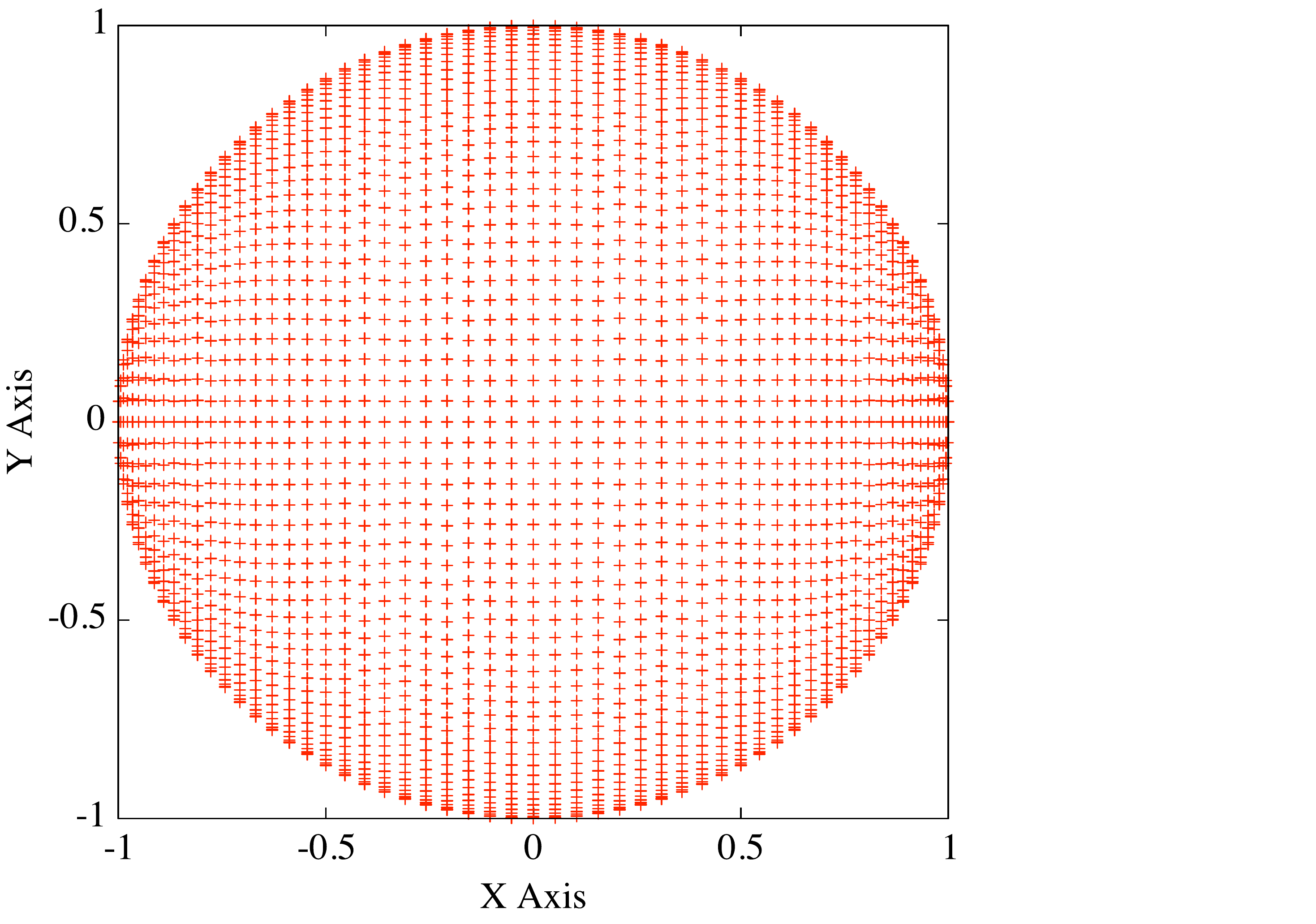}
      \caption{The final $(x_i,y_{ij})$ far field cartesian 
      coordinate system, with $N_{Xfar}=61$.}
      \label{2dy_xy_plot}
\end{figure}

\begin{figure}[h!]
  \centering
    \includegraphics[width=0.6\textwidth]{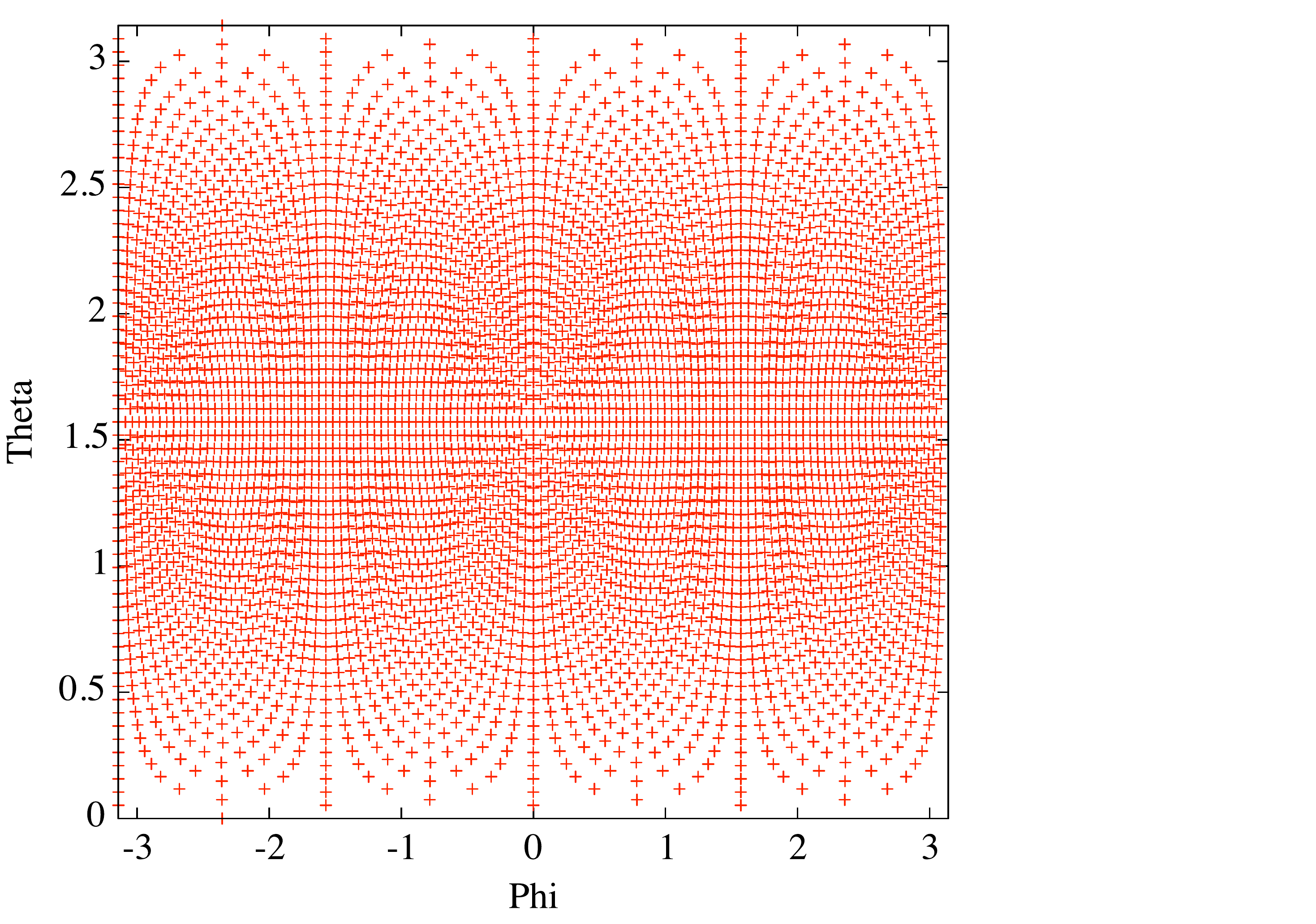}
      \caption{Sampled $(\theta,\phi)$ points on the sphere, using the 
      $(x_i,y_{ij})$ system, with $N_{Xfar}=61$. The sphere is now 
      evenly sampled, with a $\sin{\theta}$ density of sample points.}
      \label{2dy_thphi_plot}
\end{figure}

The $O(N^3)$ method amounts to calculating (\ref{N_eq}) and (\ref{L_eq}) 
more efficiently. Consider the expansion (\ref{r_phase}), but rewrite using 
cartesian coordinates for both the current surface $\bold{r}'$ and 
the 2-sphere surface $\bold{r}$:
\be
|\bold{r}-\bold{r}'| \sim r - \frac{1}{r}(xx' + yy' + zz')
\ee
The $e^{ -ik |r'| \cos{\psi} } $ term in (\ref{N_eq}) and (\ref{L_eq}) can thus be replaced 
by $e^{ -ik xx'}e^{ -ik yy'}e^{ -ik zz'}$, where we have simplified slightly by assuming 
normalized $x,y,$ and $z$: $x^2 + y^2 + z^2 = 1$.
We will also specify the dimensions of the current extraction surface $S$: let $x'_1 \le x' \le x'_2$ 
(with $N_X$ sample points), $y'_1 \le y' \le y'_2$, and $z'_1 \le z' \le z'_2$
(with $N_Y$ and $N_Z$ points respectively).
Then consider just the integration of the $X$-$Y$ plane located at $z'=z'_2$:
\be
\bold{N}(x,y) = \int_{y'_1}^{y'_2} \int_{x'_1}^{x'_2} \bold{J}_s(x',y') e^{ -ik xx'}e^{ -ik yy'}e^{ -ik zz'_2} dx'dy'
\ee
We are now mapping the spherical $\bold{N}$ surface in terms of the cartesian 
coordinates $x$ and $y$, with the $z$ position implicit in terms of the other two:
$z = \pm\sqrt{1-x^2-y^2}$.
This double integral is now separable if we define an intermediate variable 
(with 2D storage requirements) which we call $\bold{T}$. We can perform 
the $x'$ integration first (free to choose in general):
\be
\bold{T}(x,y') = \int_{x'_1}^{x'_2} \bold{J}_s(x',y') e^{ -ik xx'} dx' , \label{T_def}
\ee
and then promptly use $\bold{T}$ during the $y'$ integration to 
recover $\bold{N}$:
\be
\bold{N}(x,y) = e^{ -ik zz'_2 } \int_{y'_1}^{y'_2} \bold{T}(x,y') e^{ -ik yy'} dy' \label{N_fast}
\ee
Note that we have moved the $e^{ -ik zz'_2 }$ term outside of the integral, 
and in general we will need two copies of $\bold{N}(x,y)$: one where we 
have multiplied by $e^{ -ik z'_2 \sqrt{1-x^2-y^2}}$ (for the ``northern hemisphere")
and the other by $e^{ ik z'_2 \sqrt{1-x^2-y^2}}$ (for the ``southern").

The far field coordinates $x$ and $y$ need to be discretized as well 
(into $N_{Xfar}$ and $N_{Yfar}$ points), 
and we will see later that it is reasonable to choose $N_{Xfar}, N_{Yfar} \sim N_{\theta}, N_{\phi}$.
We have thus recovered $\bold{N}$ (that is, the portion due to the $X$-$Y$ plane at $z'$=$z'_2$)
in $O(N_{Xfar}N_Y N_X) + O(N_{Xfar}N_{Yfar} N_Y)$ operations 
(for (\ref{T_def}) and (\ref{N_fast}) respectively). 

The next $X$-$Y$ plane at $z'$=$z'_1$ naturally follows in the same fashion, but we 
have some choice in integrating $X$-$Z$ and $Y$-$Z$ planes.
In general one could still map $\bold{N}$ using the 
$x$ and $y$ coordinates to integrate the $Y$-$Z$ plane, with the analog of 
(\ref{N_fast}) becoming:
\be
\bold{N}(x,y) = e^{ -ik xx'_2 } \int_{z'_1}^{z'_2} \bold{T}(y,z') e^{ -ik z'\sqrt{1-x^2-y^2}} dy' \label{N_fast2} .
\ee
However, we will later introduce a ``2D" $y$ coordinate, which would necessitate an 
internal $y$ interpolation inside (\ref{N_fast2}).
Additionally we will later interpolate our cartesian-mapped far field 
data back into the standard spherical $\theta$,$\phi$ coordinate 
system so that it can be used by other software 
(and this interpolation process scales as $O(N^2\log{N})$, 
and therefore is not a bottleneck).
We thus choose to interpolate the $\bold{J}_s$ current $Y$-$Z$ plane to a separate 
$y,z$ far field plane: $\bold{N}(y,z)$, and likewise 
the $X$-$Z$ currents to $\bold{N}(x,z)$.
Using distinct far field planes allows the same code for (\ref{T_def}) and (\ref{N_fast}) 
to be reused after a fast internal coordinate swap.
These three far field planes can then be interpolated and summed to the same 
final $\bold{N}(\theta,\phi)$ result (and likewise for $\bold{L}(\theta,\phi)$).

Next we more closely consider our far field $(x,y)$ coordinate system
(with the additional $(y,z)$ and $(z,x)$ planes following a similar logic).
The easiest to implement are discretized uniform 
distributions (in psuedocode):

\begin{algorithm}
\caption{Uniform X}
\label{uniform_x}
\begin{algorithmic}
\State$\Delta x \gets 2/(N_{Xfar} - 1)$
\For{$i = 0$ to $(N_{Xfar} - 1)$} 
	\State$x[ i ] \gets -1 + i \Delta x$ 
\EndFor
\end{algorithmic}
\end{algorithm}

\noindent
and likewise for $y$. This does work, and provides ``ok" results.
One drawback of this system is that it wastes sample points: 
a fraction $(4-\pi)/4$ of the $(x,y)$ pairs will fall outside of the 
$x^2 + y^2 \le 1$ region and thus will not be used in the final result.
The primary drawback however is that these coordinates only sparsely 
sample the $z \sim 0$ ``equatorial" region of the 2-sphere.
This can be see in Fig. (\ref{uniform_xy_plot}), where the $(\theta,\phi)$ positions of the 
$(x,y); x^2+y^2 \le 1$ pairs have been plotted.
Increasing the resolution of $x$ and $y$ will generally drive more points 
into the equatorial region, but at a slow rate: the resolution of the cartesian coordinates 
need to be quadrupled in order to get nearby points twice as close 
to the $\theta=90$ degrees equator.

An initial improvement is to use cartesian coordinates with non-uniform 
spacing. We recommend using Chebyshev spacing (see e.g. \cite{Boyd_book}):

\begin{algorithm}
\caption{Chebyshev X}
\label{chebyshev_x}
\begin{algorithmic}
\State$\Delta \phi \gets \pi / (N_{Xfar} - 1)$
\For{$i = 0$ to $(N_{Xfar} - 1)$} 
	\State$x[ i ] \gets -\cos( i \Delta \phi )$ 
\EndFor
\end{algorithmic}
\end{algorithm}

\noindent
and likewise for $y$.
With these coordinates every $x$ and $y$ line will sample the $z=0$ equator.
This provides a more regular mapping of the sphere than ALGO[\ref{uniform_x}], 
but has the drawback that it still sparsely samples the $z \sim 0$ region for 
$\phi$ values centered around 45, 135, 225, and 315 degrees.

Our final coordinate system still uses Chebyshev coordinates, but uses 
a ``2D" $y$ coordinate system. That is, the ``1D" $x$ grid points are still 
spaced according to ALGO[\ref{chebyshev_x}], but for each $x$ sample point 
$x_i$ we create a separate line of $y$ sample points $y_{ij}$, where the 
$y$-line extends from $-\sqrt{1-x^2_i}$ to $\sqrt{1-x^2_i}$. We also scale the number 
of sample points along these $y$-lines by its relative length:
$N_{Yi} \sim N_{Xfar} \sqrt{1-x^2_i}$.
In pseudocode we have:

\begin{algorithm}
\caption{ 2D Chebyshev Y }
\label{chebyshev_y2d}
\begin{algorithmic}
\For {$i = 0$ to $(N_{Xfar} - 1)$} 
	\State$yRange \gets sqrt(1.0-x[i]*x[i])$
	\State$N_{Yi} \gets N_{Xfar} * yRange$
	\State$\Delta \phi \gets \pi / ( N_{Yi} - 1 )$
	\For{$j = 0$ to $(N_{Yi} - 1)$} 
		\State$y[ i ][ j ] \gets - yRange * \cos( j \Delta \phi )$ 
	\EndFor
\EndFor
\end{algorithmic}
\end{algorithm}

A plot of the $x_i$ and $y_{ij}$ sample locations can be seen in Fig. (\ref{2dy_xy_plot}), 
where we use $N_{Xfar} = 61$.
The corresponding plot of sampled $(\theta,\phi)$ locations is shown in 
Fig. (\ref{2dy_thphi_plot}).
Testing shows that the ALGO[\ref{chebyshev_x}] \& ALGO[\ref{chebyshev_y2d}] 
coordinate system samples the 2-sphere with $\sin{\theta}$
density (and is independent of $\phi$), which is ideal.
As a bonus, there are no longer any ``wasted" sample points: 
all $x_i$, $y_{ij}$ pairs (with $z=\pm\sqrt{1-x^2_i-y^2_{ij}}$ )
correspond to locations on the sphere.
Note that we are now restricted in our evaluation order of 
(\ref{T_def}) and (\ref{N_fast}): 
we now need to perform the $x'$ integration first (whereas we were 
free to choose before), at least if we want to avoid internal interpolation.
Note also that one is free to construct an analogous 1D $y$ and 2D $x$ 
coordinate system, where $y'$ would be integrated first. 

\begin{figure}[h!]
  \centering
    \includegraphics[width=0.6\textwidth]{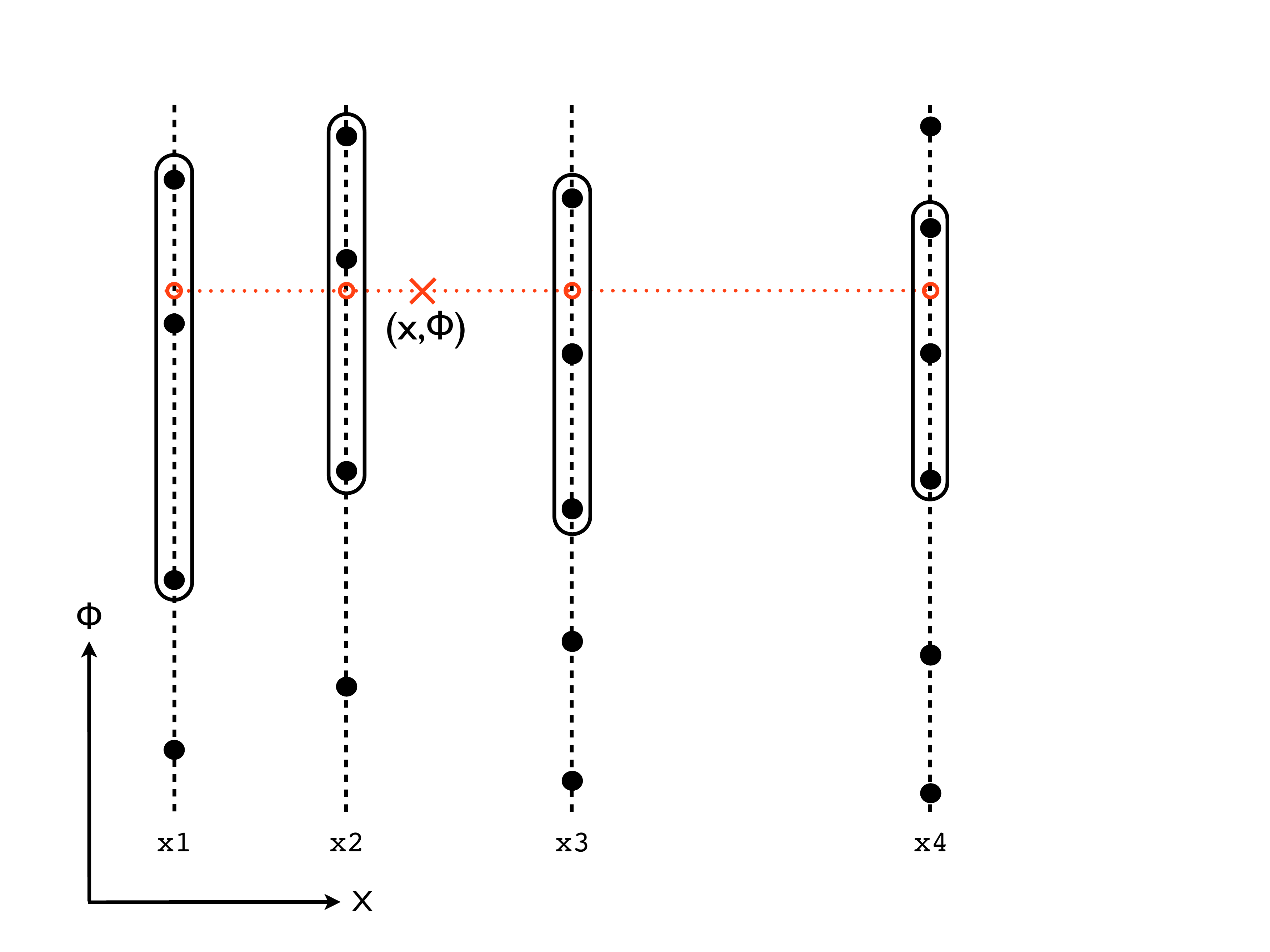}
      \caption{2D Lagrangian interpolation strategy for our cartesian coordinate system. Lagrangian interpolation 
      is first performed in the $\hat{\phi}$ direction for several nearby $\tilde{\phi}_{ij}$ data sets.
      The results from this first set of interpolations (red circles) are then used in a second 
      Lagrangian interpolation in the $\hat{x}$ direction. In this example we use quadratic interpolation 
      in the $\phi$ direction, and cubic in $x$.}
      \label{2d_lag_1_plot}
\end{figure}

\begin{figure}[h!]
  \centering
    \includegraphics[width=0.6\textwidth]{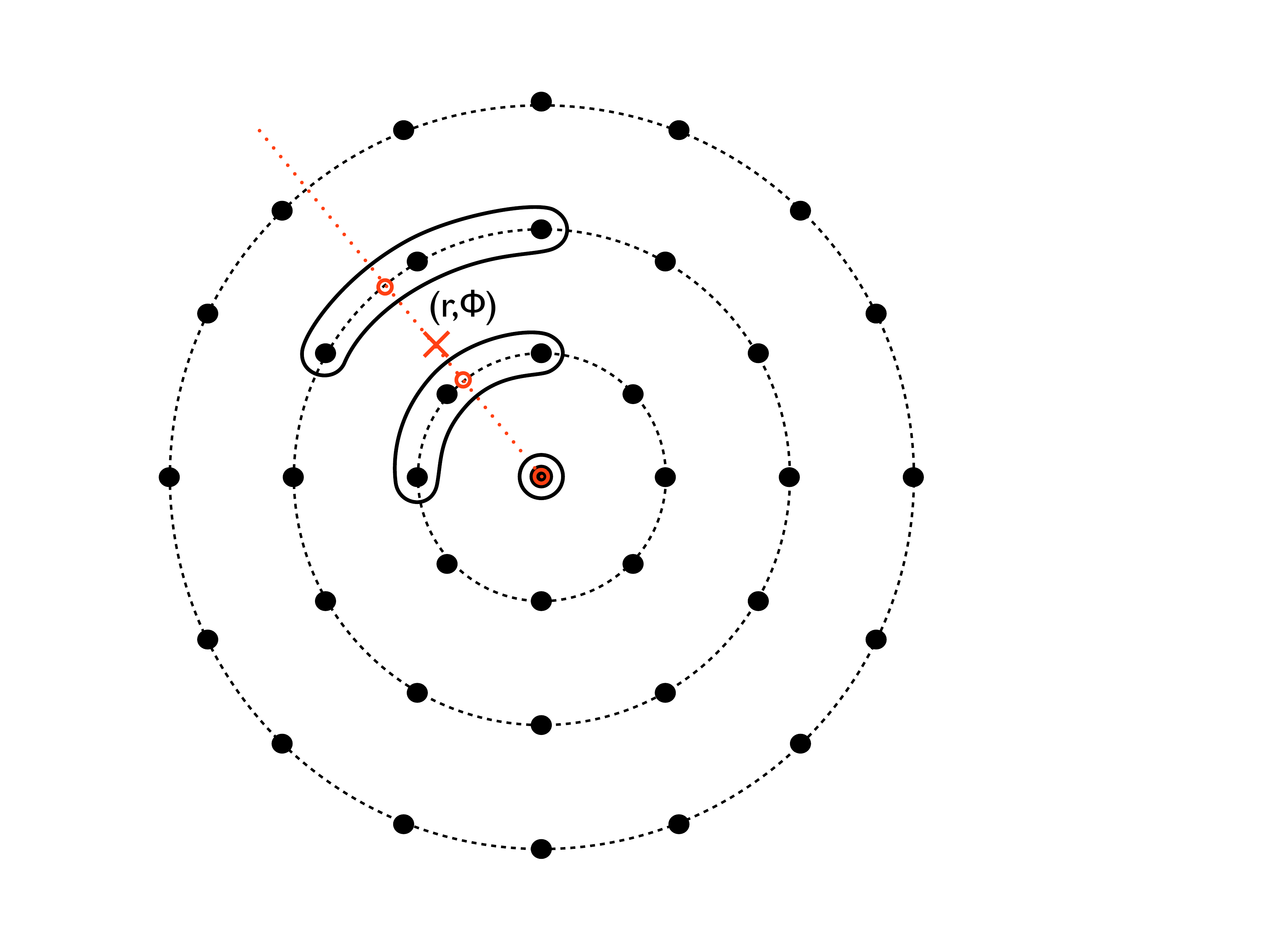}
      \caption{Polar 2D Lagrangian interpolation strategy used near the $x \sim \pm1$ end points.
      Here we switch from parameterizing positions by $x$ to $r=\sqrt{y^2+z^2}$.
      Interpolation is first performed in the $\hat{\phi}$ direction at several radii, 
      and then in the $\hat{r}$ direction.}
      \label{2d_lag_2_plot}
\end{figure}

The ability to interpolate data from one set of spherical coordinates 
to another is desirable in general, 
and it is particularly needed here since we integrate the 
currents on the $X$-$Y$, $Y$-$Z$, and $Z$-$X$ near field planes 
to separate, and respectively parallel far field planes.
We thus need to interpolate and sum the separate $\bold{N}(x,y)$, 
$\bold{N}(y,z)$ and $\bold{N}(z,x)$ data to the same $\bold{N}(\theta,\phi)$
grid to get our final answer.
We have developed a 2D Lagrangian interpolation system 
to this end.

Consider again the distribution of $(x_i,y_{ij})$ sample points as seen in Fig. (\ref{2dy_xy_plot}).
When we multiply by the 2 $z$ components (i.e. $e^{\pm ikzz'_2 }$ in (\ref{N_fast})), 
the initially planar $\bold{N}(x,y)$ data becomes 3-dimensional.
For a given $x_i$ location we view the corresponding line of $y_{ij}$ sample 
points as being promoted to set of azimuthal sample points via: $\tilde{\phi}_{ij} = \arctan{}(z_{ij}/y_{ij})$.
Thus consider interpolation to a particular $(\theta_s,\phi_s)$ location.
We first convert this to cartesian coordinates: $x_s = \sin{\theta_s}\cos{\phi_s}$, 
$y_s = \sin{\theta_s}\sin{\phi_s}$, and $z_s = \cos{\theta_s}$.
We then use a binary search to find the closest $x_i$ to $x_s$, 
and then a second binary search to find the closest $\tilde{\phi}_{ij}$ to $\tilde{\phi}_s=\arctan{}(z_s/y_s)$.
If a closest neighbor interpolation strategy sufficed, we could then directly sum the data at 
$x_i, \tilde{\phi}_{ij}$ to the field at $(\theta_s,\phi_s)$.

Much better results can be generated by using Lagrangian interpolation however 
(see e.g. \cite{Boyd_book}). In general given a set of $N$ sample points from an unknown 
function: $(x_i,f_i); \; 1\le i \le N$, one can generate interpolating function $f(x)$: 
\be
f(x) = \sum_i f_i P_i(x), \quad P_i(x) = \prod_{j, j\ne i} \frac{x-x_j}{x_i-x_j} \label{lag_interp}
\ee 
Note that Lagrangian interpolation can be dangerous to use in the general case, 
due to Runge phenomenon, but is not a problem here as we do not go to high 
order, and only utilize the interpolation near the center of a range.
2D Lagrangian interpolation can then be built on top of the 1D version, 
usually by assuming that the 2 coordinates lie along a rectilinear grid.
That is not the case for our problem, as the number and location of 
$\tilde{\phi}$ sample points varies for each $x_i$ location. However, since 
all of the $\tilde{\phi}$ sample points do share the same $x$ position for 
each azimuthal ring, we can still construct an effective ``2 pass"
Lagrangian interpolation method.

Given a point $(x_s,\tilde{\phi}_s)$ that we wish interpolate to, 
we first find the closest neighbor in 
the $x$ (i.e. $x_i$) and $\tilde{\phi}$ directions using binary searches as before. 
We then perform quartic interpolations 
to the $\tilde{\phi}_s$ position using the azimuthal data at each of the 5 
$x$ locations: $x_{i-2}, \; x_{i-1}, \; x_{i}, \; x_{i+1}, \; x_{i+2}$ .
We then use these intermediate results to perform 
a subsequent quartic interpolation in the $x$ direction to $x_s$. 
This two-pass interpolation scheme is shown in Fig. (\ref{2d_lag_1_plot}).
In general one is free to set the interpolation order, we settled on quartic in each 
direction due to the rapid convergence it provides as the resolution is increased.  
Note that the azimuthal sample points have a branch cut from $+\pi$ to $-\pi$, 
and we thus pad our azimuthal data by several sample points. For instance 
the pad point at $+\pi+\Delta \tilde{\phi}$ is set to have the same field value as 
$-\pi+\Delta \tilde{\phi}$.

We tweak the interpolation scheme to improve accuracy 
near the ends of the $x$ range, where $x \sim \pm 1$.
Here the Chebyshev nodes become closely spaced and so 
we switch from parameterizing in $x$ to $r = \sqrt{y^2+z^2}$, 
resulting in a polar coordinate system.
Interpolation first takes place in the $\tilde{\phi}$ direction, 
as before, followed by a radial interpolation.
This variation on the interpolation method is diagrammed in Fig. (\ref{2d_lag_2_plot}).

\section{Results}

The $O(N^3)$ method was inspired through work on an existing 
NTFF code base. This simplified development of the algorithm, 
as effectively only the code for (\ref{N_eq}) and (\ref{L_eq})
needed to be replaced with (\ref{T_def}) and (\ref{N_fast}) 
(along with the associated Chebyshev coordinates and interpolation tools). 

An example run of the algorithm is shown in Fig. (\ref{result_1_plot}), where we plot the directivity pattern
as found by the $O(N^3)$ method for 
a simple Vlasov antenna 
(which is formed by cutting a circular 
waveguide at an angle, resulting in a hypodermic needle shape).
The shape of the aperture causes the 
waves exit at an angle from the axis of the 
waveguide, and thus the peak directivity is offset from $\theta=0$ --
in this case with the TM01 mode the peak occurs at $\theta=25$ degrees.
The plot looks identical to the eye as compared to the same plot 
generated by the traditional $O(N^4)$ NTFF method.

\pagebreak
\begin{widetext}

\begin{figure}[h!]
  \centering
    \includegraphics[width=1.0\textwidth]{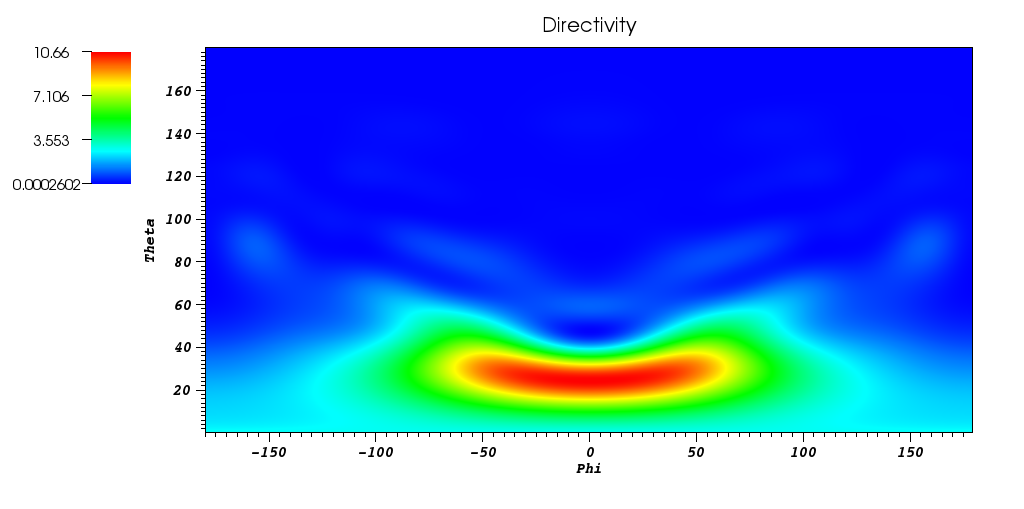}
      \caption{Directivity plot for the TM01 mode of a Vlasov antenna 
      as generated by the $O(N^3)$ NTFF method. The same plot from the standard 
      $O(N^4)$ method looks identical to the eye.}
      \label{result_1_plot}
\end{figure}

\vspace{20mm}

\begin{figure}[h!]
  \centering
    \includegraphics[width=1.0\textwidth]{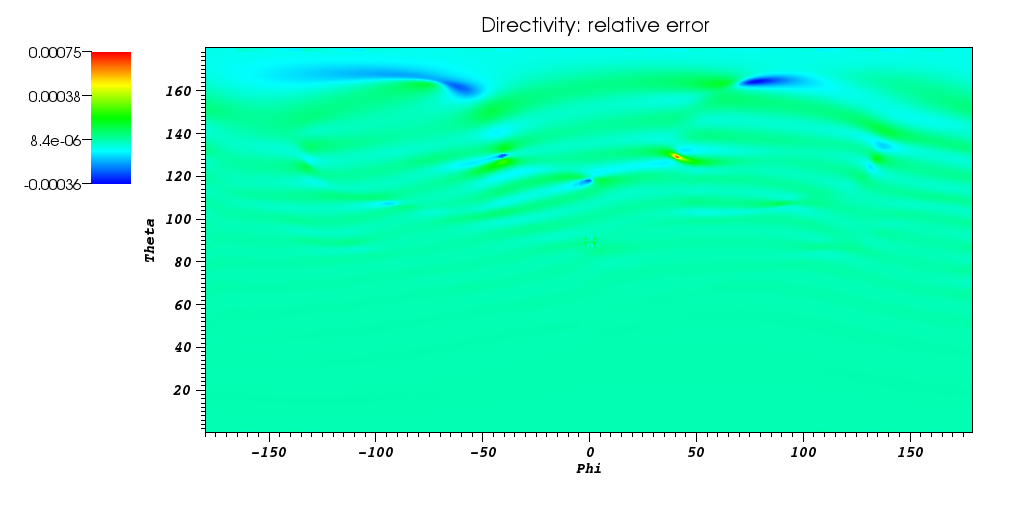}
      \caption{Relative error for the Vlasov antenna directivity between the $O(N^3)$ and $O(N^4)$ methods.
      The largest relative errors overall ($\sim 0.05 \%$) occur where the 
      directivity has the lowest values ($\sim 0.00026$). 
      Near the peaks of the directivity ($\theta \sim 25$ degrees) the largest relative errors 
      are at roughly one part in a million.}
      \label{result_2_plot}
\end{figure}

\end{widetext}

The slight differences in the directivity $D$ 
produced by the 2 methods can be seen in Fig. (\ref{result_2_plot}) where 
we plot relative 
error $(D[N^3] - D[N^4])/D[N^4]$.
The largest relative errors, on the order of $\sim 0.05 \%$ 
occur where the directivity has the lowest amplitudes, 
roughly $\sim 0.0003$, and most of the overall visible error 
occurs within the ``shadowed" region $\theta \sim 90$ 
degrees to $\theta = 180$, opposite of the orientation of the antenna. 

This error is not caused by the interpolation process: we use $N_{Xfar} = 180$
for this simulation, and one can see interpolation artifacts if 
this value is lowered, but the pattern in Fig. (\ref{result_2_plot}) remains constant 
as we go to higher values.
Instead we suspect that the relative error between the two methods 
is the fault of the $O(N^4)$ method, i.e. we expect the 
$O(N^3)$ pattern is closer to the true pattern a physical device would produce.
This is due a commonly seen phenomena \cite{Ishimaru_1978}: the backscatter pattern for 
strongly forwardly scattering objects is often poorly recovered by the 
standard NTFF method, as it involves the cancellation of large numbers 
of individual current contributions, which is subject to roundoff error.
As the $O(N^3)$ method simply involves far fewer calculations, we expect that 
truncation errors have less of a chance to accumulate, and the answer 
in the backscatter region (and between lobes in general) 
should be more accurate.
In general when it is the backscatter fields that are of interest 
(as in \cite{Backman_2000, Kim_2003,Roy_2004})
a useful technique is to remove the forward scattering near field surface 
from the far field integration sum \cite{Li_2005}. We are interested to see if the 
$O(N^3)$ method can further improve the accuracy of these results.

In general the $O(N^3)$ method correctly recovers the same far field patterns 
as found by the traditional method, and we do find that it does so much 
more quickly, as expected by the scaling analysis. 
On large FD-TD simulations with 1 degree spacing of the far fields 
($N_{\theta} = 180$, $N_{\phi} = 360$) we have generated speedups 
by a factor of $\sim 100$ and are interested in possible further optimization.
This can make a very 
big difference for large problems (such as those run on large clusters), 
when one desires the full bistatic pattern.
Where far fields patterns at a single frequency may have had to suffice previously, 
we can now do frequency sweeps. We also note that the algorithm should 
parallelize well: in a distributed environment processors that own a subset of 
the near field surface can calculate their contribution to the far field pattern, 
with all of these contributions then summed through a MPI Reduce type operation. 

\begin{acknowledgments}
The author thanks Will Dicharry, John McIver, Stephen McCracken and Shane Stafford 
for their work on an existing NTFF codebase which this project builds upon, 
and the Air Force Research Laboratory and High Performance Computing
Modernization Office for funding.
\end{acknowledgments}


\end{document}